\documentclass[journal]{IEEEtran}

\ifCLASSINFOpdf
\else
   \usepackage[dvips]{graphicx}
\fi
\usepackage{url}

\usepackage{graphicx, float,verbatim}
\usepackage{algorithmic, amsfonts, gensymb}
\usepackage{amsmath, amssymb}
\usepackage{cite, array, multirow}
\usepackage{setspace,bbding}

\begin{document}

\title{Towards Unified All-Neural Beamforming for \\ Time and Frequency Domain Speech Separation}

\author{Rongzhi Gu, Shi-Xiong Zhang, Yuexian Zou,~\IEEEmembership{Senior Member,~IEEE}, Dong Yu,~\IEEEmembership{Fellow,~IEEE}}

\markboth{}
{Shell \MakeLowercase{\textit{et al.}}: }
\maketitle

\begin{abstract}
Recently, frequency domain all-neural beamforming methods have achieved remarkable progress for multichannel speech separation. In parallel, the integration of time domain network structure and beamforming also gains significant attention. This study proposes a novel all-neural beamforming method in time domain and makes an attempt to unify the all-neural beamforming pipelines for time domain and frequency domain multichannel speech separation. The proposed model consists of two modules: separation and beamforming. Both modules perform temporal-spectral-spatial modeling and are trained from end-to-end using a joint loss function. The novelty of this study lies in two folds. Firstly, a time domain directional feature conditioned on the direction of the target speaker is proposed, which can be jointly optimized within the time domain architecture to enhance target signal estimation.  
Secondly, an all-neural beamforming network in time domain is designed to refine the pre-separated results. This module features with parametric time-variant beamforming coefficient estimation, without explicitly following the derivation of optimal filters that may lead to an upper bound. 
The proposed method is evaluated on simulated reverberant overlapped speech data derived from the AISHELL-1 corpus. Experimental results demonstrate significant performance improvements over frequency domain state-of-the-arts, ideal magnitude masks and existing time domain neural beamforming methods. 
 
\end{abstract}

\begin{IEEEkeywords}
Multichannel speech separation, all-neural beamforming, neural beamforming, time domain, end-to-end
\end{IEEEkeywords}

\IEEEpeerreviewmaketitle

\section{Introduction}

\IEEEPARstart{T}{he} rise of deep learning technology has enabled significant progress in the study of close-talk speech separation, but 
the performance of far-field speech separation in the presence of interfering speech, reverberation and noise is still far from satisfactory. To this end, microphone array based multichannel speech separation (MC-SS) models, which leverage spatial information to distinguish spatially distributed sources \cite{vincent2014blind}, are more favorable in the current state-of-the-art distant-talking enhancement and recognition systems \cite{pfeifenberger2015multi,tu2017design,wang2020complex}.

Mainstream MC-SS methods formulate speech separation as a supervised learning problem in time-frequency (T-F) domain or time domain. For T-F domain approaches, the raw mixture is transformed to complex-valued T-F representations using short-time Fourier transform (STFT). The final objective is to approach the target speech spectrogram with T-F mask estimation. The spatial information can be injected into the separation process, either via the feature formulation \cite{gu2020enhancing,han2021multi,luo2019fasnet} or network structures designed to operate across microphone channels \cite{luo2020end,pandey2022multichannel}. For time domain approaches, the raw mixture is directly input to the separation network, aiming to learn a mapping from the mixture signal to the target signal. 

To fully utilize the spatial information in the multichannel signal, beamforming techniques are extensively adopted for multichannel signal processing. 
In early works \cite{du2016ustc,heymann2016neural,erdogan2016improved,higuchi2016robust}, beamforming is integrated with the neural network based mask estimator, the process of which is irrelevant to the training of the mask estimator. Recently, the beamforming algorithm is implemented as a fully differentiable network and optimized jointly with the mask estimation neural network \cite{heymann2017beamnet,ochiai2017unified,boeddeker2017computation,xu2020neural}, referred as \emph{neural beamforming}. This neural beamforming framework has demonstrated superior performance over the previous two-stage framework \cite{zhang2020adl}. More recently, the idea of \emph{all-neural beamforming} in T-F domain is explored \cite{li2021mimo,xu2021generalized}. Compared to neural beamforming that implements well-designed beamforming algorithms with network layers and operations, all-neural beamforming methods do not adopt the closed-form solutions derived from certain optimization criteria (such as MMSE and Max-SNR) and constraints. Instead, the solution of all-neural beamforming is learned completely data-driven towards the target signal.

Despite the success achieved by the all-neural beamforming in T-F domain, there are still some limitations that have not been addressed. As analyzed and experimentally reported in \cite{heitkaemper2020demystifying}, the complex nature of STFT, as well as the limited time resolution of computed spectrograms, causes an obvious upper bound to the separation performance. Motivated by recent explorations of neural beamforming in time domain \cite{luo2022time}, this study proposes to establish an all-neural beamforming network in time domain for multichannel speech separation.

The contributions of this work are as followed:
\begin{itemize}
    \item We propose a novel directional feature in time domain that explicitly conditions on the direction-of-arrival (DOA) of the target speaker. This feature can be simply yet efficiently integrated into other advanced time domain MC-SS methods as a target DOA cue to further enhance the separation performance. To the best of our knowledge, this is the first work to perform target speech separation in time domain based on DOA information.
    
    \item We propose an all-neural beamforming network in time domain, which leverages the pre-separated waveforms to learn parametric filter-and-sum beamforming coefficient estimation directly in time domain. The learned filters are response-invariant for broadband speech processing and yield the same beam pattern for different frequencies for preventing potential signal distortion. 

    \item We make an attempt to unify the all-neural beamforming framework for time and frequency domain multichannel target speech separation. The proposed methods, as well as the aforementioned popular methods, both in frequency and time domain, are contrastively compared and analyzed on a simulated reverberant dataset derived from the AISHELL-1 corpus. 
\end{itemize}

The rest of the paper is organized as follows. Section \ref{sec:related_works} reviews the related works. Section \ref{sec:mch_tss} presents the overall framework of the proposed MC-SS method, which consists of two stages, respectively elaborated in Section \ref{sec:sep_stage} and Section \ref{sec:bf_stage}. Section \ref{sec:exp} and Section \ref{sec:rlt} describe the experimental setup and analyze the results. Section \ref{sec:conclusion} concludes the paper. 

\section{Related Works}
\label{sec:related_works}
This section reviews related works among three aspects of MC-SS: temporal-spectral-spatial feature formulation, neural beamforming in frequency domain and time domain. 
\vspace{-0.1cm}
\subsection{Temporal-spectral-spatial feature formulation}

Many efforts have been made to leverage the temporal-spectral-spatial information contained in multichannel signals \cite{lianwu2019multi,wang2018multi,luo2021implicit,han2021multi,gu2020enhancing,koyama2020efficient}. From the input feature perspective, interaural phase difference (IPD) is well defined in the T-F domain based on STFT \cite{allen1977short}, which manifests the time delay conditioned on the DOA. 
When the target DOA is given, which can be estimated via speaker localization techniques, a type of directional feature \cite{wang2018spatial,wang2019combining,gu2019neural} is empirically designed in T-F domain. This type of directional feature indicates T-F bins that are more likely to be dominated by the target speech by comparing the similarity of the observed IPDs with theoretical target IPDs (T-IPDs) conditioned on target DOA. The abovementioned spatial features are hand-crafted and computed based on STFT. 
The complex nature of STFT brings an extra burden for the model to learn the correlations between magnitude and phase components \cite{luo2019convtasnet}. 
Another potential limitation is that to meet the short-time stationary hypothesis of STFT, a window length of 25-32ms is usually adopted, which constrains the time resolution of the spectrogram and is therefore hard to satisfy the W-disjoint orthogonality hypothesis \cite{rickard2002approximate}. 

Owing to the abovementioned limitations of STFT, a recent trend is to extract spatial information via time domain architectures in an end-to-end optimization manner \cite{gu2020enhancing,luo2020end,han2021multi,koyama2020efficient,zhang2020end,zhang2021time}. Inspired by the state-of-the-art single-channel Conv-TasNet \cite{luo2019convtasnet}, which replaces STFT and inverse STFT (iSTFT) with an encoder-decoder structure, \cite{gu2020enhancing} proposes to use multiple sets of learnable filters to transform the multichannel mixture into multichannel feature maps with convolution operation. Based on this architecture, \cite{gu2020enhancing,zhang2020end} derive interchannel convolution difference (ICD) features that implicitly characterize the spatial difference between sources. Similarly, decorrelation on two channels of feature maps has been performed to capture the interchannel differential spatial information \cite{han2021multi}. Apart from pairwise processing, a transform-average-concatenate (TAC) module, which aggregates the transformed features of all channels to make global decisions, is proposed in \cite{luo2020end}. 

Although learning spatial features with time domain network architectures has shown promising results, it is difficult to leverage the DOA information within these architectures. The reason lies in the fact that the learnable filters do not hold the clear physical definition as STFT, thereby causing difficulty in computing the theoretical values conditioned on the target DOA. Also, the aforementioned methods all follow the reference-channel mask estimation framework to perform speech separation, which potentially brings nonlinear distortion and causes performance degradation. 

\vspace{-0.1cm}
\subsection{Neural beamforming in frequency domain}

As the potential nonlinear distortion brought by masking based separation methods, also, to better utilize the multichannel signals at the output side, beamforming is more favorable in mainstream MC-SS methods \cite{heymann2016neural,higuchi2016robust}. Owing to better generalization ability and computation efficiency, frequency domain beamforming is more frequently used in these works. 

MC-SS methods that combine deep neural network (DNN) with well-studied beamforming techniques can be mainly divided into three categories:
\begin{itemize}
    \item \emph{T-F masking based beamforming} \cite{erdogan2016improved,higuchi2016robust,heymann2016neural,du2016ustc}. In these methods, the DNN is served as a target mask or signal estimator, trained to produce a more precise estimation to the supervision signal. The estimated mask is then used to calculate the signal statistics for subsequent beamforming algorithms based on different optimization criteria. Since the beamforming stage is irrelevant to the mask estimation stage, this kind of two-stage framework may cause sub-optimal performance. 
    \item \emph{Neural beamforming} \cite{heymann2017beamnet,ochiai2017unified,boeddeker2017computation,xu2020neural,zhang2020adl,wang2021sequential}. This method formalizes the closed-form solution of the beamforming algorithm as a fully differentiable network, and jointly optimized with the mask estimation network with a unified objective towards the target signal. Through joint optimization, the mask estimation is optimized for the beamforming purpose. 
    However, the beamforming solution still follows specific statistical optimization criterion, which brings an obvious performance upper bound.
    \item \emph{All-neural beamforming} \cite{ochiai2017unified,xiao2016deep,meng2017deep,he2020spatial,zhang2022all,zhang2020adl}. This method directly estimates the beamforming weights with DNNs. Xiao et al. \cite{xiao2016deep} have firstly attempted to estimate the complex-valued filter-and-sum beamformer. Further, \cite{ochiai2017unified} explores time-variant filter estimation for better adapting to the dynamic acoustic environment. 
    Very recently, Zhang et. al. \cite{zhang2020adl} propose an all-neural beamforming method, which integrates mask estimation and beamforming weight estimation into a unified network that can be trained from end-to-end.
    The beamforming coefficient estimation is parametric and no longer follows the closed-form solutions of the traditional beamformers, only optimized with the final speech separation objective. 
\end{itemize}

Except for T-F masking, the integration of complex spectral mapping and adaptive beamforming has also shown its effectiveness \cite{wang2020complex}. Besides the aforementioned methods that are fully formulated in T-F domain, motivated by the superior performance of single-channel time domain speech separation network (e.g., TasNet \cite{luo2019convtasnet}), methods that combine single-channel TasNet with frequency domain beamforming have also been explored \cite{ochiai2020beam,chen2021beam}. 
In this way, more accurate signal statistics are produced by TasNet for the beamforming algorithm to approach the performance upper bound of frequency domain beamforming. 

Despite the great progress achieved with all-neural beamforming (AN-BF) methods in T-F domain, the aforementioned limitations of STFT may still hinder the separation performance, including the balance between time and frequency resolution and difficulty for real-valued DNNs to deal with complex-valued features and operations. 
\vspace{-0.3cm}
\subsection{Neural beamforming in time domain}

Considering the limitations of separation in frequency domain, and motivated by the recent single-channel speech separation state-of-the-arts in time domain \cite{luo2019convtasnet,luo2020dprnn,subakan2021attention}, a recent research trend formulates MC-SS in time domain or latent domain, where the latent domain is formed by a set of learnable filters \cite{gu2020enhancing}. 

As an early trial for robust automatic speech recognition (ASR), \cite{li16f_interspeech,sainath2017multichannel} propose a unified architecture to model the fine time structure from the raw waveforms. The network architecture is featured with time domain learnable filters spanning across the microphone channels. 
Recently, \cite{luo2019fasnet} proposes a filter-and-sum network (FaSNet) for learning adaptive time domain beamforming weights from normalized cross-correlation features. More lately, \cite{luo2021implicit} alternates the filter-and-sum beamforming operation in the latent domain formulated by the learnable filters. Very recently, \cite{luo2022time} proposes an iterative framework that performs sequential latent domain speech separation and time domain multichannel Wiener filtering. 

Motivated by the great potentials demonstrated by neural beamforming in time domain, this work develops a fully time domain all-neural beamforming (AN-BF) network for MC-SS. To process the broadband speech, existing AN-BF in T-F domain performs sub-band decomposition and derives narrowband beamformers independently at each frequency \cite{zhang2020adl,xu2021generalized,li2022taylorbeamformer}. As an alternative, we propose to learn a response-invariant broadband finite-impulse response (FIR) filter directly in time domain for yielding the same beam pattern for different frequencies. This property prevents the potential signal distortion \cite{benesty2008minimum,chen2008microphone} for the downstream ASR task.

\section{Multichannel Target Speech Separation}
\label{sec:mch_tss}
\begin{figure}[b]
    \centerline{\includegraphics[width=7.5cm]{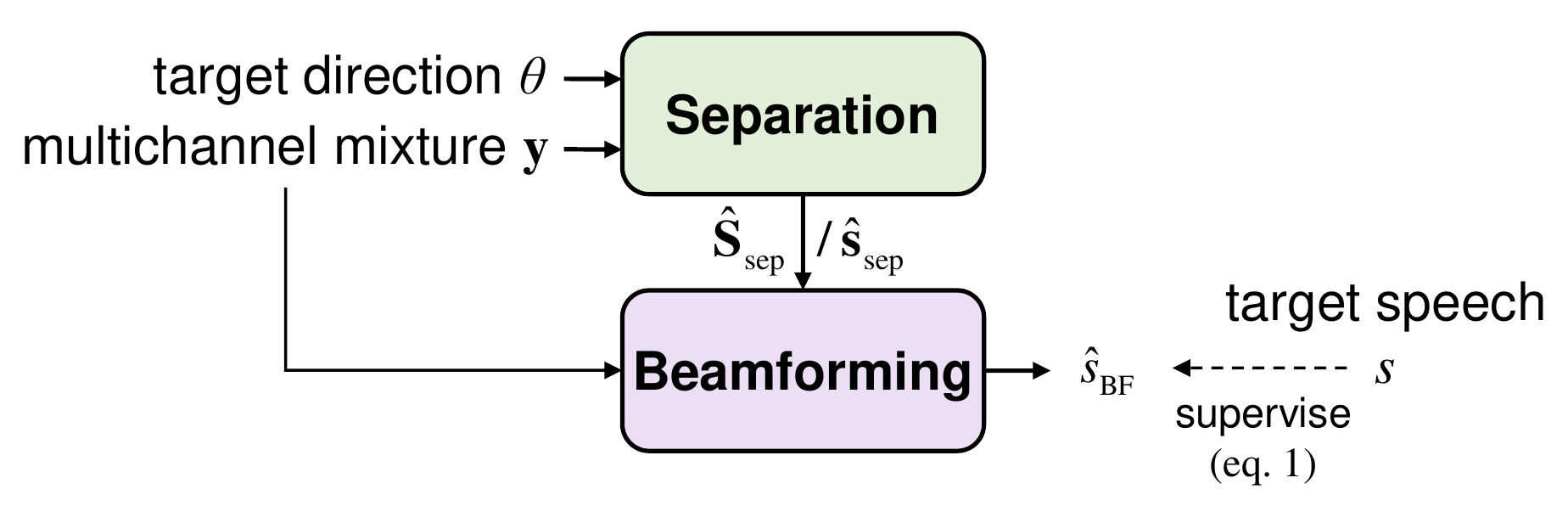}}
    \caption{The proposed multichannel target speech separation framework, which is consist of separation stage and beamforming stage.}
    \label{fig:sec2}
\end{figure}

This paper aims to extract the target speech $s$ at the reference channel from $M$-channel mixture signal $\mathbf{y}$, utilizing the given DOA $\theta$ of the target speaker to facilitate MC-SS.  

As shown in Fig. \ref{fig:sec2}, the proposed MC-SS framework is consist of two stages. The first stage is separation, which utilizes a DNN to estimate the complex spectrogram of the target speech $\hat{S}_\text{sep}$ or time domain target speech $\hat{s}_\text{sep}$. The predicted spectrograms or signals at the separation stage can be viewed as a coarse estimation of the target signal, which will be refined by the following beamforming stage. Based on the pre-separated results, the beamforming stage aims to estimate beamforming coefficients for capturing the speech coming from the target direction, while suppressing the interference from other directions. The final output of the proposed MC-SS model is $\hat{s}_\textit{BF}$.

\begin{figure*}[t]
\centerline{\includegraphics[width=14cm]{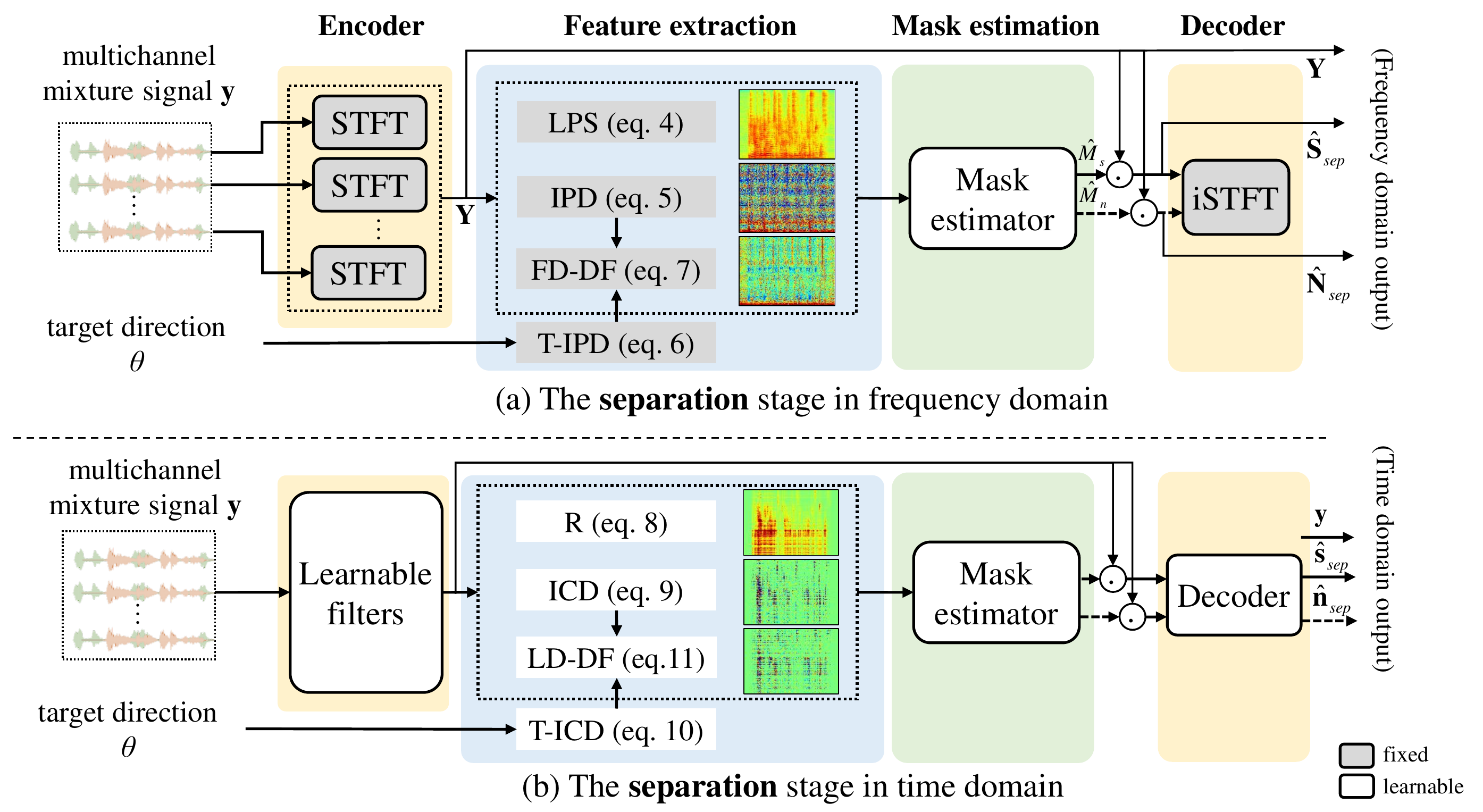}}
\caption{The \textbf{separation} stage in frequency domain and time domain for multichannel speech separation methods. The separation stage consists of 3 modules: encoder, feature extraction and mask estimation. }
\label{fig:separation_stage}
\end{figure*}

The entire framework is trained from end-to-end under the supervised learning paradigm, using target speech $s$ as the supervision signal. Following the common practice \cite{bahmaninezhad2019comprehensive}, scale-invariant signal-to-distortion ratio (SI-SDR) \cite{le2019sdr} is adopted as the loss function:
\begin{equation}
\left\{
\begin{array}{lr}
s_{\text{target}}:=\frac{\left<\hat{s}_\textit{BF}, s\right>s}{\left\|s\right\|_{2}^{2}} \\
e_{\text{noise}}:=\hat{s}_\textit{BF}-s_{\text{target}}  \\
\text{SI-SDR}:=10\log_{10}\frac
{\left\|s_{\text{target}}\right\|_{2}^{2}}
{\left\|e_{\text{noise}}\right\|_{2}^{2}}
\end{array}
\right.
\label{eq:si_sdr}
\end{equation}
SI-SDR is a loss function defined in the logarithmic time domain that measures the ratio between the target signal and the error signal, which also has a close relation to the logarithmic mean square error (MSE) criterion \cite{heitkaemper2020demystifying}.

\section{Separation Stage}
\label{sec:sep_stage}
This section will firstly describe the separation stage in frequency domain in Section \ref{subsec:fd_sep}, the design of which is built upon our previous work \cite{gu2019neural}. Then, the separation stage in time domain is presented in Section \ref{subsec:td_sep}, where we propose a method for jointly learning temporal-spectral-spatial features from time domain mixture signals.  
\vspace{-0.1cm}
\subsection{Speech separation in frequency domain}
\label{subsec:fd_sep}
\subsubsection{Overview}

As shown in Figure \ref{fig:separation_stage} (a), the separation stage in frequency domain consists of 3 modules: 1) STFT, which decomposes the multichannel mixture signal on a set of fixed complex filters; 2) Feature extraction, which aims to fully leverage the temporal-spectral-spatial information to extract effective cues for mask estimation;  3) T-F mask estimator, aiming to estimate a complex ratio mask (cRM) \cite{williamson2015complex} for the target speech at the reference channel based on the extracted features, as well as for the interfering speech. The output of the separation stage is the estimated multichannel spectrograms for the target speech $\hat{\mathbf{S}}_\text{sep}$ and residual sound $\hat{\mathbf{N}}_\text{sep}$, and the complex spectrogram $\mathbf{Y}$ of the multichannel mixture.

\subsubsection{Feature extraction with STFT}

The STFT operation can be viewed as convolving the signal with FIR exponential filters windowed by $w[n]$, as shown in Figure \ref{fig:filter} (a). These exponential filters can be combined as a complex convolution kernel $\mathbf{F} \in \mathbb{C}^{N\times F}$ that is parameterized by:
\begin{equation} 
    \mathbf{F}_{nf} = w[n]\mathbf{e}^{-j\frac{2\pi}{N}nf}
\label{eq:stft_kernel}
\end{equation}
where $n$ denotes the sample index within the window with length $N$, $f$ denotes the frequency band index, $F$ is the number of total frequency bands. Define the $f$-th filter as $\mathbf{F}_f=[\mathbf{F}_{1f},...,\mathbf{F}_{Nf}]^{\mathsf{T}}$, and define the segmentation of time domain signal with window length $N$ and hop size $H$ as  $\overline{\mathbf{y}}(t)=\left[ \mathbf{y}(tH),...,\mathbf{y}(tH+N-1) \right] ^\mathsf{T} \in \mathbb{R}^{M\times N}$, then the STFT of the $m$-th channel of the mixture signal is calculated by convolving $\mathbf{y}^m$ with $\mathbf{F}$:
\vspace{-0.15cm}
\begin{equation} 
\mathbf{Y}^m(t,f)= \overline{\mathbf{y}}^m(t)\circledast\mathbf{F}_f = \sum^{N-1}_{n=0} \mathbf{y}^m(tH+n)\mathbf{F}_{nf}
\label{eq:tf_stft}
\end{equation}
where $\circledast$ denotes the convolution operation, $\mathbf{Y}$ is the multichannel mixture spectrogram, $t$ is the frame index. 
Towards the learning target of the cRM, the input features are designed to manifest the differences between simultaneous speech both in the spectral and spatial domains. First, the logarithm power spectra (LPS) of the mixture signal at the reference channel is calculated as the spectral feature:
\begin{equation} 
\text{LPS}_{t,f}=20\lg\left ( \left |\overline{\mathbf{y}}^\text{ref}(t)\circledast\mathbf{F}_f \right |\right)
\label{eq:tf_spectral}
\end{equation}
where `$\text{ref}$' is the reference microphone index. Except for spectro-temporal properties, spatial diversity is a significant cue for separating spatially distributed speech sources. Therefore, IPD between each microphone pair $p$ is extracted to manifest the spatial difference between sources:
\begin{equation} 
\text{IPD}^{(p)}_{t,f}=\angle(\overline{\mathbf{y}}^{p_1}(t)\circledast\mathbf{F}_f)-\angle(\overline{\mathbf{y}}^{p_2}(t)\circledast\mathbf{F}_f)
\label{eq:tf_spatial}
\end{equation}
IPDs can be viewed as observed values that contain both target and interference components. To extract components that only relate to the target speech, we employ the target DOA information to construct the theoretical IPD between each microphone pair. Target IPD (T-IPD) is calculated by the phase difference that a unit impulse coming from $\theta$ will experience at the $p$-th microphone pair:
\begin{equation} 
\text{T-IPD}^{(p)}_{f}(\theta)=\angle(\delta[n]\circledast\mathbf{F}_f)-\angle(\delta[n-\tau^{(p)}(\theta)]\circledast\mathbf{F}_f)
\label{eq:tf_tpd}
\end{equation}
where $\delta[n]$ is the unit impulse function, and $\tau^{(p)}(\theta)= d^{(p)}\cos \theta f_s/c$ is the pure delay experienced by the planar wave impinging from $\theta$ at the $p$-th microphone pair, where $d^{(p)}$ is the distance between the $p$-th microphone pair, $c$ is the sound velocity, $f_s$ is the sampling rate.
Notably, the T-IPD is commonly computed by $\text{T-IPD}_f^{(p)}(\theta)= 2\pi f \tau^{(p)}(\theta)$. To determine the dominance of the target speech from $\theta$, the cosine similarity between the observed IPDs and theoretical T-IPDs is measured at each T-F bin \cite{chen2018multi}, named as frequency domain directional feature (FD-DF):
\begin{equation} 
\text{FD-DF}_{t,f}(\theta)={\sum}_p\left <
\text{IPD}^{(p)}_{t,f},\text{T-IPD}^{(p)}_f(\theta)
\right > 
\label{eq:tf_df}
\end{equation}

\subsubsection{Mask Estimation}

The calculated LPS, IPDs between multiple microphone pairs and FD-DF are concatenated along the frequency axis to form a joint feature as the input to the T-F mask estimator. Conv-TasNet is served as the backbone structure of the T-F mask estimator, and it can be replaced by other state-of-the-art network structures, such as DPRNN \cite{luo2020dprnn}, SepFormer \cite{subakan2021attention}, DUNet \cite{hu2020dccrn}, etc. The output layers of the separation network are two point-wise convolution layers, respectively produce the estimated cRMs of the target speech and the interfering speech, i.e., $\hat{M}_s \in \mathbb{C}^{T\times F}$ and $\hat{M}_n\in \mathbb{C}^{T\times F}$. The estimated masks are then used to compute the separated spectrograms $\hat{\mathbf{S}}_\text{sep}=\hat{M}_s \circ \mathbf{Y} \in \mathbb{C}^{M\times T\times F}$ and $\hat{\mathbf{N}}_\text{sep}=\hat{M}_n \circ \mathbf{Y}\in \mathbb{C}^{M\times T\times F}$. Note that the mask is shared across all the signal channels to apply to $\mathbf{Y}$.

\vspace{-0.3cm}
\subsection{Speech Separation in Time Domain}
\label{subsec:td_sep}

As discussed in the introduction, owing to the complex nature of features extracted in the frequency domain, it is difficult for a real-valued network without exquisite design to model the relations between magnitude and phase parts \cite{yin2020phasen}. In addition, the center frequency of STFT filters is distributed with uniform intervals, which limits the model to learn from a linear frequency domain. Furthermore, to meet the short-time stationary hypothesis, the constraint on the window length (about 25-32 ms) induces a relatively low time resolution of the computed features, which could be a significant reason for the limited separation performance analyzed in \cite{heitkaemper2020demystifying}. 

Inspired by recent advances on time domain neural architectures for MC-SS \cite{gu2020enhancing, luo2020end, luo2021implicit, koyama2020efficient}, we propose to learn a real-valued feature space, which is formulated by learnable filters, for better representing the temporal-spectral-spatial characteristics of the multichannel mixture signal. 

\subsubsection{Overview}
As shown in Figure \ref{fig:separation_stage} (b), the separation stage in time domain consists of 4 modules: 1) Learnable filters are served as the waveform encoder to transform the multichannel mixture signals into a multichannel representation in latent domain; 2) Feature learning, which extracts temporal-spectral-spatial features in the latent space; 3) Mask estimator that estimates the masks of the target and interfering speech in the latent domain based on the computed features; 4) Decoder that converts the masked mixture representation back to time domain signal. The output of separation stage is the estimated signals $\hat{\mathbf{s}}_\text{sep}$, $\hat{\mathbf{n}}_\text{sep}$, and the multichannel mixture signal $\mathbf{y}$.

\subsubsection{Feature learning with learnable filters}
\begin{figure}[t]
\centerline{\includegraphics[width=8.5cm]{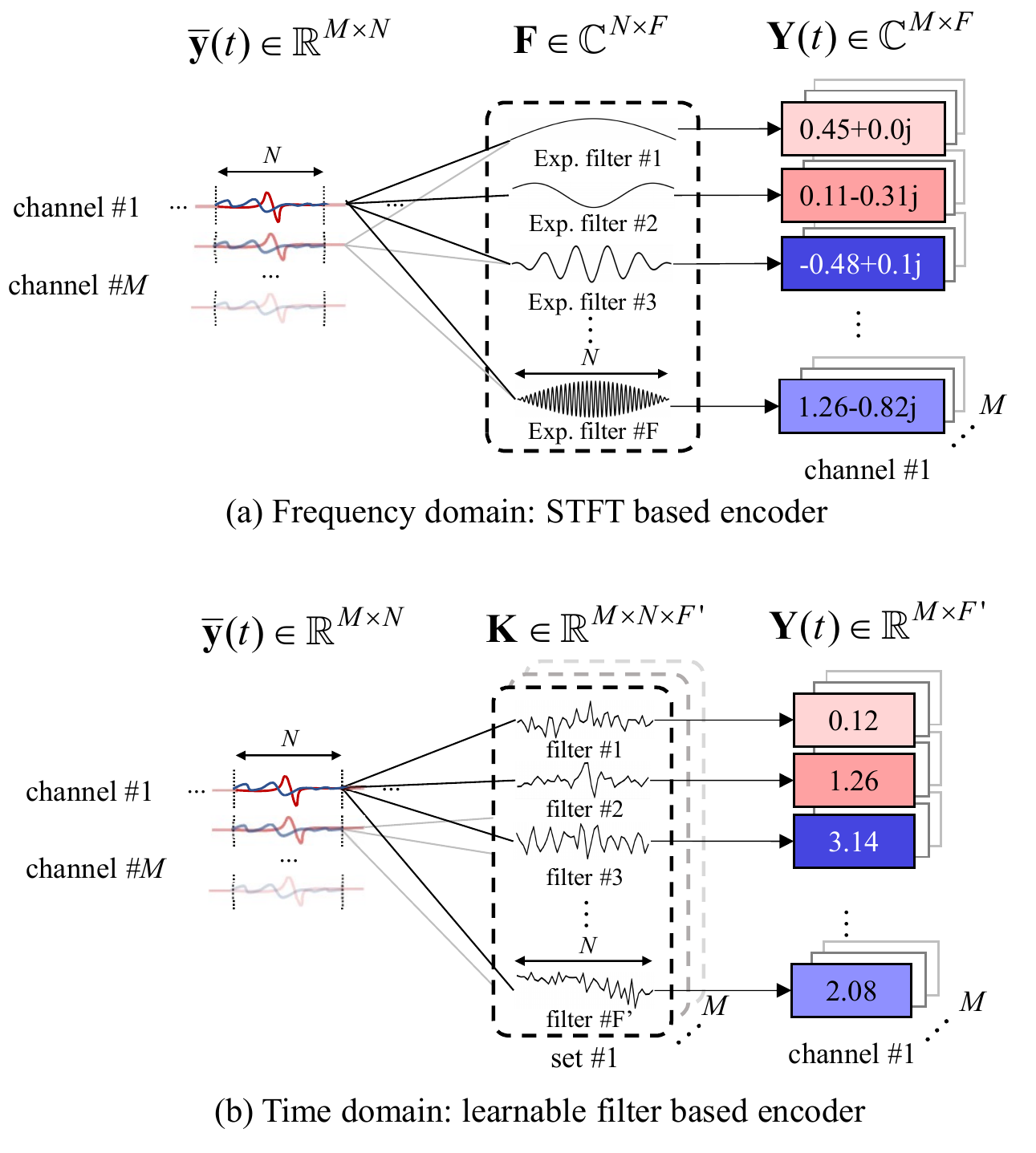}}
\caption{The illustration of transforming the $t$-th frame of multichannel mixture $\overline{\mathbf{y}}(t)$ with (a) STFT based encoder $\mathbf{F} \in \mathbb{C}^{N \times F}$; (b) learnable filters based encoder $\mathbf{K} \in \mathbb{R}^{M\times N \times F'}$. $N$ is the window length.}
\label{fig:filter}
\end{figure}

Instead of STFT, we explore $M$ sets of learnable nonorthogonal real-valued filters $\mathbf{K}\in \mathbb{R} ^{M \times N \times F'}$ to transform each short segment of the multichannel mixture $\overline{\mathbf{y}}(t) \in \mathbb{R} ^ {M \times N}$ to a multichannel representation $\mathbf{Y}(t) \in \mathbb{R} ^ {M \times F'}$, as shown in Figure \ref{fig:filter} (b), where $F'$ is the number of filters, $N$ is the frame size. Following our previous work \cite{gu2020enhancing}, to ensure mapping each channel to the same feature space, each set of filters is associated with a set of reference filters $\mathbf{K}^0 \in \mathbb{R} ^{N \times F'}$, and varies with a learnable window function $w^m \in \mathbb{R}^{N\times 1}$: $\mathbf{K}^{m} = w^m \mathbf{K}^0$. By shifting the window and integrating $M$ channels, a multichannel representation that contains rich temporal-spectral-spatial information could be computed. 

First, the mixture representation at the reference channel is considered as the spectral feature \cite{luo2019convtasnet}, which is the convolution product between $\overline{\mathbf{y}}^{\text{ref}}$ and the reference filter $\mathbf{K}^0$:
\begin{equation} 
\text{R}_{t,f'}=\sigma(\overline{\mathbf{y}}^{\text{ref}}(t)\circledast\mathbf{K}^0_{f'})
\label{eq:learn_spectral}
\end{equation}
where $\sigma$ is the activation function, $f'$ is the filter index. Inspired by the IPD formulation, the interchannel convolution differences (ICDs)  \cite{gu2020enhancing} between two channels of the feature maps are computed, implicitly manifesting the spatial difference between sources:
\begin{equation} 
\text{ICD}^{(p)}_{t,f'}=(\overline{\mathbf{y}}^{p_1}(t)\circledast\mathbf{K}^{p_1}_{f'})-(\overline{\mathbf{y}}^{p_2}(t)\circledast\mathbf{K}^{p_2}_{f'})
\label{eq:learn_spatial}
\end{equation}
From Eq. \ref{eq:learn_spatial}, ICD is defined in a feature space formulated by learnable filters $\mathbf{K}$, which are updated at every iteration and do not hold the physical definition as STFT. 

Conventionally, the DOA is used to form the steering vector \cite{higuchi2016robust}, determine the active T-F bins \cite{araki2007blind}, compute the second-order statistics of the target speech \cite{thiergart2013informed}, etc. Most of these usages are designed for frequency domain methods and are difficult to integrate into the time domain approaches. To this end, we propose to explicitly leverage the target DOA within the time domain architecture for better extracting the target speech. From a signal processing perspective, each set of filters $\mathbf{K}^m$ can be viewed as a linear time-invariant (LTI) system, whose impulse response is defined as its output for a unit impulse input. For a microphone pair ($p_1, p_2$), we input a unit impulse to the system corresponding to $p_1$, and a delayed unit impulse to the system corresponding to $p_2$. The expected response difference conditioned on the target DOA $\theta$, named target ICD (T-ICD) is then defined as the output difference between these two systems to capture the time delay:
\begin{equation} 
\text{T-ICD}^{(p)}_{f'}(\theta)=\delta[n]\circledast\mathbf{K}^{p_1}_{f'}-\delta[n-\tau^{(p)}(\theta)]\circledast\mathbf{K}^{p_2}_{f'}
\label{eq:learn_tcd}
\end{equation}
Comparing Eqs. \ref{eq:learn_spatial} and \ref{eq:learn_tcd}, if ($t,f'$) is dominated by the target speech from $\theta$, the similarity between $\text{ICD}^{(p)}$ and the target location related $\text{T-ICD}^{(p)}$ will be large; otherwise, it will be small. Following this concept, the directional feature in latent domain (LD-DF) is defined as follows:
\begin{equation} 
\text{LD-DF}_{t,f'}(\theta)={\sum}_{p} \left <
\text{ICD}^{(p)}_{t,f'},\text{T-ICD}^{(p)}_{f'}(\theta)
\right > 
\label{eq:learn_df}
\end{equation}

\subsubsection{Mask Estimation}

The computed R (Eq. \ref{eq:learn_spectral}), ICDs (Eq. \ref{eq:learn_spatial}) and LD-DF (Eq. \ref{eq:learn_df}) are concatenated along the feature axis and input to the mask estimation network to estimate the masks for the target and interfering speech. Finally, the masked multichannel mixture representation will be converted back to the time domain signal $\hat{\mathbf{s}}_\text{sep}$ and $\hat{\mathbf{n}}_\text{sep}$ by the decoder, respectively. The decoder is a 1D deconvolution layer \cite{luo2019convtasnet} with the same kernel size and stride as the waveform encoder. 

\vspace{-0.1cm}
\section{Beamforming Stage}
\label{sec:bf_stage}
This section will present the designs of the beamforming stage in frequency and time domain, respectively. First, in Section \ref{subsec:fd_anbf}, motivated by the impressive improvements achieved by AN-MVDR \cite{zhang2020adl}, the beamforming stage in frequency domain combines the AN-BF designs in \cite{zhang2020adl} and \cite{xu2021generalized}, to further enhance the separation performance while reducing the nonlinear distortion brought by single-channel masking. Also, the framework is extended with different beamforming formulations. Then, in Section \ref{subsec:td_anbf}, inspired by the success of AN-BF in frequency domain and recent explorations on time domain neural beamforming \cite{luo2022time}, we design an AN-BF network in time domain to establish a completely end-to-end MC-SS framework.

\begin{figure}[b]
\centerline{\includegraphics[width=\linewidth]{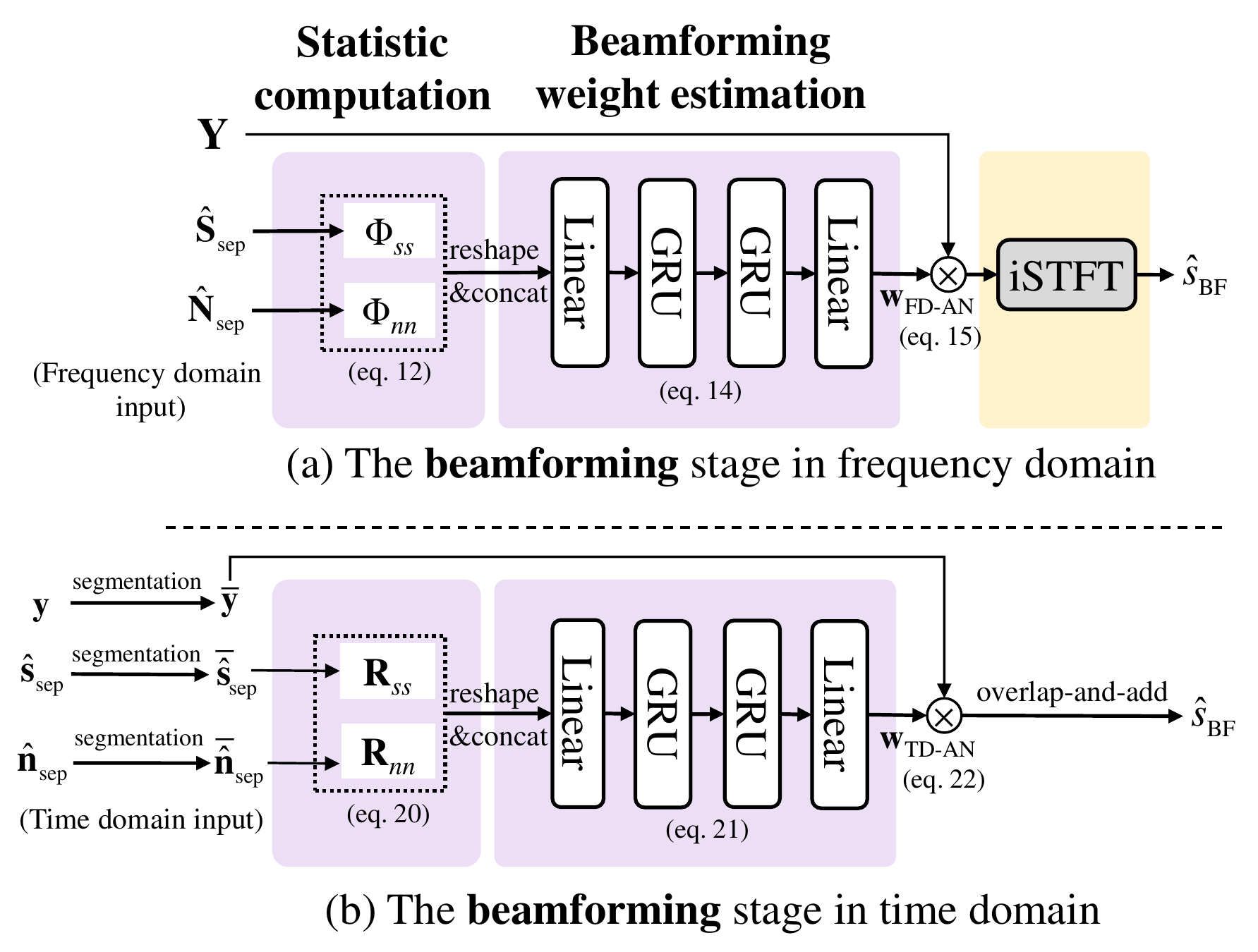}}
\caption{The \textbf{beamforming} stage in frequency domain and time domain for multichannel speech separation. The beamforming stage consists of 3 modules: signal statistic computation, beamforming weight estimation and beamforming. }
\label{fig:beamforming_stage}
\end{figure}
\vspace{-0.2cm}
\subsection{All-neural beamforming in frequency domain}
\label{subsec:fd_anbf}
\subsubsection{Method}
As shown in Figure \ref{fig:beamforming_stage} (a), all-neural beamforming leverages the estimated spectrograms to compute second-order signal statistics for estimating the weights of a filter-and-sum beamformer.
To be specific, firstly, the target and interference spatial correlation matrices (SCMs) are calculated:
\begin{equation}
    \begin{split}
     \Phi_{ss}(t, f) &= \hat{\mathbf{S}}_\text{sep}(t, f)\hat{\mathbf{S}}_\text{sep}^\mathsf{H}(t, f) \\
     \Phi_{nn}(t, f) &= \hat{\mathbf{N}}_\text{sep}(t, f)\hat{\mathbf{N}}_\text{sep}^\mathsf{H}(t, f) \\
    \end{split}
\label{eq:phi_ss}
\end{equation}
where $\mathsf{H}$ denotes the conjugate transpose matrix. When using batch-wise training, each complex SCM is with the shape of ${B\times T\times F\times M \times M}$, where $B$ is the batch size. AN-MVDR follows the sub-band beamforming concept, where the beamforming weights at each frequency band are separately estimated and applied. Therefore, these two SCMs, $\Phi_{ss}$ and $\Phi_{nn}$, are firstly reshaped to $\mathbb{C}\in ^{BF \times T\times M^2}$ and then, their real and imaginary parts are concatenated along the last dimension to feed into a feedforward layer with input size of $4M^2$. Next, two recurrent layers are followed to approximate to the matrix inversion and PCA operation in the closed-form solution of the time-invariant MVDR beamformer in frequency domain \cite{chen2008microphone}:
\begin{equation}
\mathbf{w}_\text{FD-eq-MVDR}(f)=\frac{\Phi_{nn}^{-1}(f) \mathbf{v}(f)}{\mathbf{v}^{\mathsf{H}}(f) \Phi_{nn}^{-1}(f) \mathbf{v}(f)}\mathbf{u}
\label{eq:eq_mvdr}
\end{equation}
where the target acoustic transfer function $\mathbf{v}(f)=PCA(\Phi_{ss}(f))$, $\Phi_{nn}(f)$ are summed over the whole input sequence to obtain more stable statistics, $\mathbf{u}\in \mathbb{R}^{M}$ is a one-hot vector that marks the reference channel. 
Instead of calculating the beamforming weights using Eq.\ref{eq:eq_mvdr}, in the all-neural beamforming method, the solution of AN-MVDR beamformer is derived by:
\begin{equation}
\begin{split}
\mathbf{w}_\text{FD-AN}(t, f) &=g \left (  \Phi(t, f) , \mathbf{h}_\text{FD-AN}(t-1, f)  \right )
\\
\Phi(t, f)&=\text{concat}(\Phi_{ss}(t,f), \Phi_{nn}(t, f) )
\end{split}
\label{eq:fd_an}
\end{equation}
where $\mathbf{w}_\text{FD-AN}\in \mathbb{C}^{B\times T\times F \times M}$, $\mathbf{h}_\text{FD-AN}(t-1,f)$ is the hidden state of the GRU at $t$-1 th time step, $g(\cdot)$ indicates the function of recurrent neural network layers to compute the frame-wise beamformer weights. 
Finally, the estimated filter-and-sum coefficients are multiplied to the multichannel mixture spectrogram to obtain the beamformed complex spectrogram:
\begin{equation}
\hat{S}_\text{FD-BF}(t,f)=\mathbf{w}_\text{FD-AN}^\mathsf{H}(t,f)\mathbf{Y}(t,f)
\label{eq:fd_an_bf}
\end{equation}
The beamformed spectrogram is then converted back to the estimated target speech $\hat{s}_\text{BF}$ with iSTFT.

\subsubsection{Different beamforming formulations}
Different from the MVDR beamformer that is deduced under target distortionless constraint, multichannel Wiener filter (MCWF) is derived without such constraint. Since the adopted loss function (Eq. \ref{eq:si_sdr}) is closely related to time domain MSE criterion, which is also the optimization criterion for MCWF, we explore all-neural MCWF beamforming within the framework. The closed-form solution of MCWF in frequency domain is as follows \cite{souden2009optimal}: 
\begin{equation}
\mathbf{w}_\text{FD-eq-MCWF}(f)=\Phi_{yy}^{-1}(f)\Phi_{ys}(f)
\label{eq:eq_mcwf}
\end{equation}
where $\Phi_{yy}(f)$ and $\Phi_{ys}(f)$ denote the self-correlation matrix of mixture spectrogram and cross-correlation matrix between the mixture and target spectrograms, respectively. Similar to the all-neural MVDR formulation, the second-order statistics in Eq. \ref{eq:fd_an} can be substituted with:
\begin{equation}
\Phi'(t, f)=\text{concat}(\Phi_{yy}(t,f), \Phi_{ys}(t, f) )
\label{eq:fd_an_mcwf_phi}
\end{equation}
where the spatial correlation matrics are computed as:
\begin{equation}
\begin{split}
\Phi_{yy}(t,f) &= \mathbf{Y}(t,f)\mathbf{Y}^\mathsf{H}(t,f) \\
\Phi_{ys}(t,f) &= \mathbf{Y}(t,f) \left( \hat{M}_s(t,f) \circ \mathbf{Y}^\text{ref}(t,f) \right)^\mathsf{H}\\
\end{split}
\label{eq:fd_an_mcwf}
\end{equation}
With the formulation of all-neural MCWF, only the mask of target speech $\hat{M}_s$ needs to be estimated at the separation stage.

\subsection{All-neural beamforming in time domain}
\label{subsec:td_anbf}

\subsubsection{Method}

As shown in Figure \ref{fig:beamforming_stage} (b), all-neural beamforming in time domain leverages the separated signals and the multichannel mixture signal to compute second-order signal statistics for estimating the weights of a time-variant beamformer in time domain. 
Conventionally, the solution of the MVDR beamformer in the time domain is derived by minimizing the MSE of the residual interference and noise with the constraint that the target signal is not distorted, as follows \cite{bai2013timeMVDR}:
\begin{equation}
\mathbf{w}_\text{TD-eq-MVDR} = \frac{\mathbf{R}_{nn}^{-1}\mathbf{h}}
{\mathbf{h}^{\mathsf{T}}\mathbf{R}_{nn}^{-1}\mathbf{h}}\mathbf{u}
\label{eq:td_mvdr}
\end{equation}
where $\mathbf{h}$ is the steering vector of the target direction, which can be derived by applying PCA to the spatial self-correlation matrix of the target speech $\mathbf{R}_{ss} = E\left[ \mathbf{s}(k)\mathbf{s}^\mathsf{T}(k) \right ]$, $\mathbf{R}_{nn} = E\left[ \mathbf{n}(k)\mathbf{n}^\mathsf{T}(k) \right ]$ is the spatial self-correlation matrix of the interference-plus-noise signal, $E[\cdot]$ denotes the mathematical expectation and $k$ is the discrete sampling stamp. 

In our implementation, we take the output from the decoder as the estimated target speech $\hat{s}_\text{sep}$ to obtain the estimation for $\mathbf{R}_{ss}$. The SCM of time domain signals are computed by:
\begin{equation}
    \begin{split}
    \mathbf{R}_{ss}(t,n) &=  \overline{\hat{\mathbf{s}}}_\text{sep}(t,n)
        \overline{\hat{\mathbf{s}}}_\text{sep}^\mathsf{T}(t,n)  \\
    \mathbf{R}_{nn}(t,n) &= \overline{\hat{\mathbf{n}}}_\text{sep}(t,n)
        \overline{\hat{\mathbf{n}}}_\text{sep}^\mathsf{T}(t,n) 
    \end{split}
\label{eq:R_ss_nn}
\end{equation}
where $\overline{\hat{\mathbf{s}}}_\text{sep}(t,n)\in \mathbb{R}^{M\times 1}$, $\overline{\hat{\mathbf{n}}}_\text{sep}(t,n)\in \mathbb{R}^{M\times 1}$ denote the $n$-th sampling point in the $t$-th segmented frame of estimated target speech and interfering speech, respectively. When using batch-wise training, each SCM ($\mathbf{R}_{ss}$ or $\mathbf{R}_{nn}$) is with the shape of ${B\times T\times N\times M \times M}$. To fully leverage the intra-segment relation, we aggregate the sequential and spatial information within the $t$-th frame of these two SCMs. Specifically, $\mathbf{R}_{ss}$ and $\mathbf{R}_{nn}$ are reshaped to ${B\times T\times NM^2}$, where $\mathbf{R}_{ss}(t)\in \mathbb{R}^{B\times NM^2}$ and $\mathbf{R}_{nn}(t)\in \mathbb{R}^{B\times NM^2}$. 

Following the AN-BF design in frequency domain, the matrix inversion as well as the PCA operation in Eq.\ref{eq:td_mvdr} is substituted with the recurrent neural network. These two SCMs are then concatenated along the last dimension to feed into a feedforward layer with input size of $2NM^2$. Next, two recurrent layers are followed to implement the matrix inversion and PCA operation in Eq. \ref{eq:td_mvdr}. 

In summary, the proposed solution of AN-BF in time domain is derived by:
\begin{equation}
\begin{split}
    \mathbf{w}_\text{TD-AN}(t) &=g' \left (  \mathbf{R}(t), \mathbf{h}_\text{TD-AN}(t-1) \right)\\
    \mathbf{R}(t)&=\text{concat} \left (\mathbf{R}_{ss}(t), \mathbf{R}_{nn}(t) \right ) 
\end{split}
\label{eq:td_an}
\end{equation}
where $g'(\cdot)$ indicates the neural network layers to compute the filter-and-sum beamforming weights in time domain. 
Finally, the estimated filter-and-sum coefficients are multiplied to the multichannel mixture signal to obtain the beamformed target signal:
\begin{equation}
\hat{s}_{\text{TD-BF}}(t,n)=\mathbf{w}_\text{TD-AN}^\mathsf{T}(t,n)\overline{\mathbf{y}}(t,n)
\label{eq:td_an_bf}
\end{equation}
The estimated target signal is then obtained using overlap-and-add method. 

\subsubsection{Different beamforming formulations}
The multichannel Wiener filter can also be applied in time domain. According to the minimum MSE (MMSE) criterion, the optimal Wiener filter in time domain is deduced as follows:
\begin{equation}
\mathbf{w}_{\text{TD-eq-MCWF}} = \mathbf{R}^{-1}_{yy}\mathbf{r}_{ys}
\label{eq:td_mcwf}
\end{equation}
where $\mathbf{R}_{yy}$ is the spaital self-correlation matrix of the multichannel mixture, $\mathbf{r}_{ys}$ is the cross-correlation vector between the mixture and the target speech. In our implementation, these second-order statistics are computed as follows:
\begin{equation}
    \begin{split}
        \mathbf{R}_{yy}(t,n) &=  \overline{\mathbf{y}}(t,n)
        \overline{\mathbf{y}}^\mathsf{T}(t,n)  \\
        \mathbf{r}_{ys}(t,n) &= \overline{\mathbf{y}}(t,n)
        \overline{\hat{s}}_\text{sep}(t,n) 
    \end{split}
\label{eq:R_ss}
\end{equation}
where $\overline{\hat{s}}_\text{sep}(t)$ only takes the reference channel of the estimated  target speech. Therefore, the spatial correlation matrices of all-neural beamforming with Wiener-like formulation in Eq. \ref{eq:td_an} can be rewritten as:
\begin{equation}
\begin{split}
\mathbf{R}'(t)&=\text{concat} \left (\mathbf{R}_{yy}(t), \mathbf{r}_{ys}(t) \right )
\end{split}
\label{eq:td_an_mcwf}
\end{equation}

\subsubsection{Multi-channel mask estimation for short window size} 
\label{subsec:mch_mask}
In frequency domain approaches, the estimated mask at the reference channel is shared across all microphone channels for computing the SCM. This is because the window size is typically set as 25-32 ms in frequency domain methods, which is much longer than the maximum time delay for a wave to travel across the microphone array, i.e., 2.35 ms for an array with 80 cm diameter. Therefore, the spectral magnitude diversity between channels can be neglected. However, for time domain approaches that have greatly benefited from the high time resolution, the estimated mask at the reference channel is no longer suitable for sharing across all the microphone channels, since the window size of 2-2.5 ms is quite close to the time delay between channels. 
Although the formulation of conventional MCWF only involves the target speech estimation at the reference channel (Eq. \ref{eq:td_mcwf}), the spatial information of the target speech will be neglected during the all-neural beamforming processing. 
Therefore, we propose to estimate multichannel masks for the target and interfering speech at the separation stage, for facilitating all-neural beamforming in time domain. Take the MCWF case as an example, the SCMs are then computed by:
\begin{equation}
\begin{split}
\mathbf{R}''(t)&=\text{concat} \left (\mathbf{R}_{yy}(t), \mathbf{R}_{ys}(t) \right ) \\
\mathbf{R}_{ys}(t,n) &= \overline{\mathbf{y}}(t,n)
\overline{\hat{\mathbf{s}}}_\text{sep}^{\mathsf{T}}(t,n) 
\end{split}
\label{eq:an_mch_mcwf}
\end{equation}
where $\overline{\hat{\mathbf{s}}}_\text{sep}(t,n)$ takes the $n$-th sampling point from the $t$-th segmented frame of the estimated multi-channel target source image signal. In the experiments, we will show the multi-channel mask estimation strategy significantly improves the speech separation performance.

\subsection{Discussion on Unification}

\begin{figure}[t]
\centerline{\includegraphics[width=\linewidth]{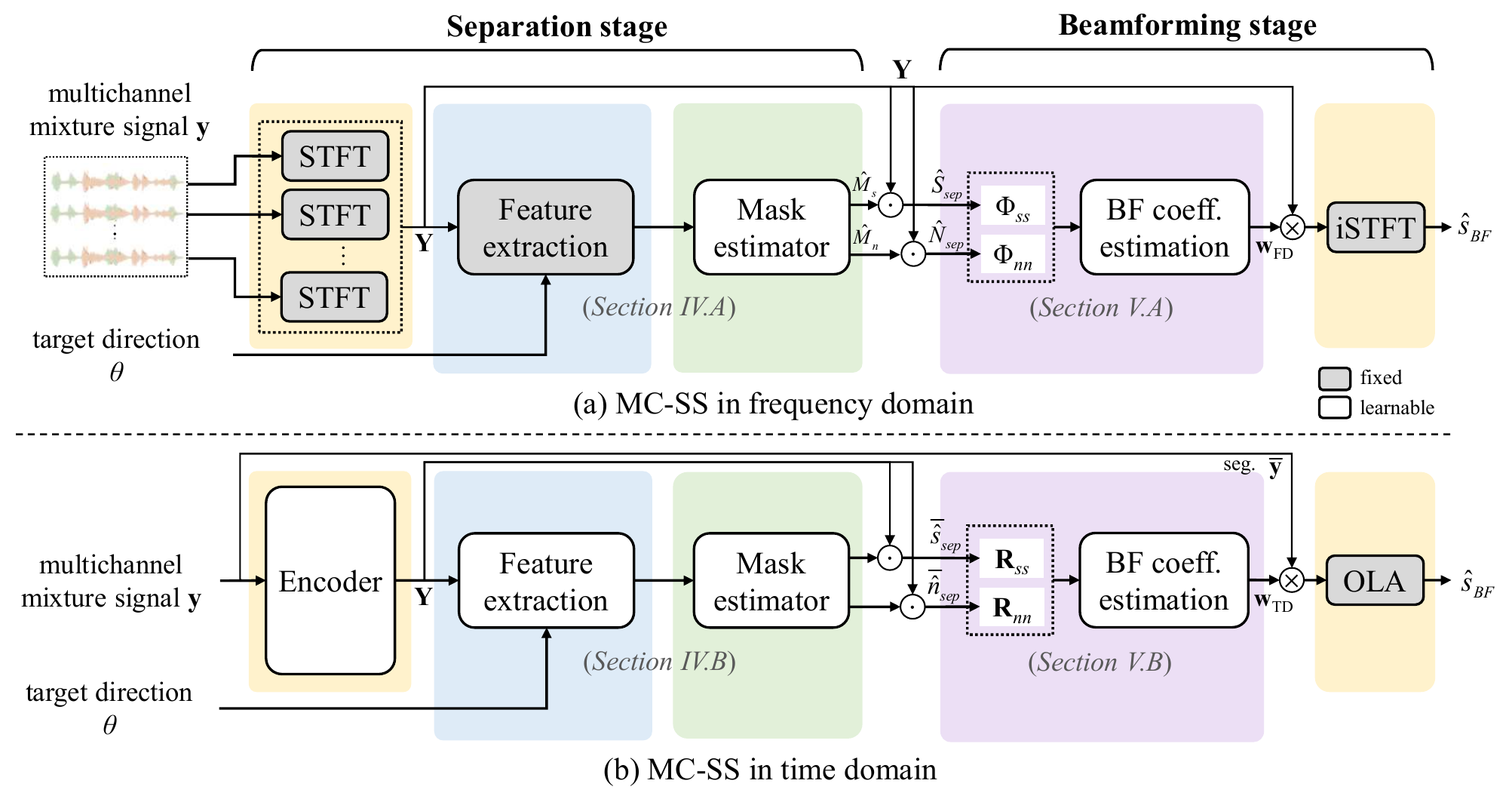}}
\caption{The unified framework for frequency domain and time domain multichannel speech separation.}
\label{fig:unified}
\end{figure}

The unification of frequency- and time-domain pipelines can be illustrated in Figure~\ref{fig:unified}. All the components, including feature design, mask estimation and beamforming operations, have the same unified form. For example, the STFT (Eq. \ref{eq:stft_kernel}) in frequency domain can be viewed as a special case of learnable filters $\mathbf{K}$ in Eq. \ref{eq:learn_spectral} in time domain. Thus, all the features in both domains such as spectral features (Eq.~\ref{eq:tf_spectral} $\rightarrow$ Eq.~\ref{eq:learn_spectral}), spatial features (Eq.~\ref{eq:tf_spatial} $\rightarrow$ Eq.~\ref{eq:learn_spatial}) and directional features (Eq.~\ref{eq:tf_df} $\rightarrow$ Eq.~\ref{eq:learn_df}) can be unified via learnable filters. 
Similarly, for frequency- and time-domain beamformings, the beamforming statistics (Eq.~\ref{eq:eq_mvdr} $\rightarrow$ Eq.~\ref{eq:td_mvdr} and Eq.~\ref{eq:eq_mcwf} $\rightarrow$ Eq.~\ref{eq:td_mcwf}) and beamforming operations (Eq.~\ref{eq:fd_an_bf} $\rightarrow$ Eq.~\ref{eq:td_an_bf}), can also be unified in the same form based on a certain Transform to the raw wave. That certain Transform in frequency domain is the Fourier Transform, and in time domain it becomes the identity Transform. 


%

\section{Experimental Setup}
\label{sec:exp}

\begin{table*}[h]
\small
  \caption{SI-SDR (dB), PESQ and CER (\%) results of MC-SS models at separation stage in frequency domain and time domain. }
  \label{tab:rlt_sep}
  \centering
  \begin{tabular}{c|l|l|cccc|c|c|c}
    \hline
    \hline
    \multirow{2}{*}{\textbf{Domain}} &
    \multirow{2}{*}{\textbf{Approach}} &
    \multirow{2}{*}{\textbf{Input Features}} &
    \multicolumn{5}{c|}{\textbf{SI-SDR (dB)}} &
    \multirow{2}{*}{\textbf{PESQ}} &
    \multirow{2}{*}{\textbf{CER (\%)}} \\ 
    & & & $<$15\degree & 15-45\degree & 45-90\degree & $>$90\degree & Avg & & \\
    \hline
    - & Mixture & - & -0.1 & 0.1 &-0.3 & 0 & -0.1 & 2.10 & 75.24\\
    - & Reverb. clean & - & $\infty$ & $\infty$ & $\infty$ & $\infty$ & $\infty$ &  4.50 & 1.74 \\
    \hline \hline
    FD & cRM & LPS & 8.97 & 8.97 & 8.99 & 9.07 & 8.99 & 2.83 & 26.55 \\
    FD & cRM & 8-ch Real+Imag & 10.66 & 11.09 & 12.04 & 12.32 & 11.43 & 2.89 & 24.79 \\    
    FD & cRM & LPS, IPD &  9.08 & 9.58 & 10.07 & 10.02 & 9.64 & 3.01 & 18.58 \\
    FD & cRM & LPS, IPD, FD-DF & 11.80 & 12.28 & 12.71 & 13.08 & 12.40 & 3.23 & 10.06\\
    \hline
    TD & target speech & R & 9.72 & 9.71 & 9.81 & 9.86 & 9.75 & 2.94 & 22.49 \\
    TD & target speech & 8-ch R & 6.88 & 8.99 & 12.46 & 14.39 & 10.18 & 2.92 & 23.73 \\
    TD & target speech & R, ICD & 8.27 & 9.71 & 12.80 & 13.86 & 10.84 & 2.97 & 16.36\\
    TD & target speech & R, ICD, LD-DF & 13.34 & 13.89 & 15.08 & 15.86 & \textbf{14.40} & \textbf{3.26} & \textbf{8.68} \\
    \hline \hline
  \end{tabular}
\end{table*}

\subsection{Dataset}
To evaluate and compare the MC-SS methods both in frequency domain and time domain, we simulate a multichannel reverberant dataset based on AISHELL-1. The speech sampling rate is 16 kHz. There are 360,000, 10,000 and 4,000 two-speaker mixtures (about 450 hours in total) for training, validation and evaluation, respectively. 
We use an 8-element ($M$=8) linear array, with spacings of 15-10-5-20-5-10-15 cm. The multichannel speech signals are generated by convolving single-channel clean signals with both simulated RIRs using the image-source method \cite{ISM}. The distance between each speaker and the center of microphone array is in the range of 0.5—6m.
The reverberation time RT60 is sampled in the range of 0.1s—0.7s. The speakers that appear in the test set do not overlap with the training set, and all speakers are assumed to not change their directions during speaking. The room size ranged from 4$\times$4$\times$2.5 to 10$\times$8$\times$6 m$^3$ (length$\times$width$\times$height). The signal-to-interference rate ranged from -6 to 6 dB.

\subsection{Features}
For the STFT analysis, we use a 32 ms ($N$=512) square root Hanning window with a hop size of 16 ms ($H$=256). IPDs and T-IPDs are extracted among six microphone pairs, (1,8), (2,7), (3,6), (4,5), (5,8), and (4,8), to sample different microphone spacings.
For the settings of learnable filters, we use a 2.5 ms ($N$=40) short window and a 1.25 ms ($H$=20) hop size. ICDs and T-ICDs are extracted using the same microphone pairs as IPDs and T-IPDs. 
Only the target's directional feature is used in all the experiments, since we assume only the target DOA is known. 
For the number of filters in each set, we set $F'$=256 according to Conv-TasNet. 

\subsection{Network structure and Training configuration}

For separation stage, Conv-TasNet \cite{luo2019convtasnet} is adopted as the backbone separation network. Conv-TasNet is consist of stacked 1-dimensional convolution blocks to increase the receptive field over the input sequence. The hyperparameters of Conv-TasNet are coincident with those in \cite{luo2019convtasnet}. 

For neural beamforming (eq-BF), after the second-order statistic computation, the beamforming weights are computed using corresponding criterion. In this case, the parameter increment is very small.

For all-neural beamforming (AN-BF), the beamforming network is the combinations of linear layers and GRU layers. 
For AN-BF in frequency domain, the input and output dimension of the first feedforward layer is $4\times 8^2$ and $180$, respectively. Then, the number of cells in two unidirectional GRU layers are $90$. The output dimension of the last linear layer are $8\times2$, which outputs the the real and imaginary parts of the frame-wise beamforming weights.
For AN-BF in time domain, the input and output dimension of the first feedforward layer is $2\times8^2$ and $32$. Then, the number of cells in two unidirectional GRU layers are $256$. The output dimension of the last linear layer is $8\times 40$, which estimates the time domain beamforming weights. 

Batch normalization \cite{ioffe2015batch} is adopted to reduce the training duration. The entire network is trained on 4-second mixture chunks using the Adam optimizer \cite{kingma2015adam} with early stopping. The learning rate is initialized as 1e$^{-3}$ and will be decayed by 0.5 when the validation loss has no improvement for three consecutive epochs. 

\subsection{Evaluation metrics}
SI-SDR, PESQ and Chinese character error rate (CER) are adopted as the evaluation metrics for the assessment of separation accuracy and speech quality. The reverberant clean target speech is used as reference for SI-SDR and PESQ computation. The ASR model \cite{tian2022integrating,tian2022consistent} is trained on 60 k hours large-scaled Mandarin speech data. It should be noted that the adopted ASR model takes the logarithm Mel-scale filter bank as the input feature, therefore the CER result would not reflect the phase estimation accuracy, the same is true of the PESQ. The separation performance is evaluated under different azimuth difference ranges between speakers, where $<15\degree$, $15-45\degree$, $45-90\degree$ and $>90\degree$ account for 27\%, 30\%, 22\% and 21\%, respectively. 

\section{Results analysis}
\label{sec:rlt}

\begin{table*}[t]
\small
  \caption{SI-SDR (dB), PESQ and CER (\%) results of multi-channel target speech separation models in frequency domain and time domain.  Note that the phase estimation accuracy can't be demonstrated by PESQ and CER metrics.}
  \label{tab:rlt_bf}
  \centering
  \begin{tabular}{c|l|c|c|cccc|c|c|c}
    \hline
    \hline
    \multirow{2}{*}{\textbf{Domain}} &
    \multirow{2}{*}{\textbf{Approach}} &
    \multirow{2}{*}{\textbf{\#param.}} &
    \textbf{MACs} &
    \multicolumn{5}{c|}{\textbf{SI-SDR (dB)}} &
    \multirow{2}{*}{\textbf{PESQ}} &
    \multirow{2}{*}{\textbf{CER (\%)}} \\ 
    & & & \textbf{(G/s)} & $<$15\degree & 15-45\degree & 45-90\degree & $>$90\degree & Avg & & \\
    \hline
    - & Mixture & - & - & -0.1 & 0.1 &-0.3 & 0 & -0.1 & 2.10 & 75.24\\
    - & Reverb. clean & - & - & $\infty$ & $\infty$ & $\infty$ & $\infty$ & $\infty$ &  4.50 & 1.74\\
    \hline \hline
    FD & eq-MVDR & 9.3M & 0.57 & 10.73 & 11.33 & 12.90 & 13.85 & 12.02 & 3.20 & 17.66\\
    FD & eq-MCWF & 9.2M & 0.56 & 10.83 & 11.45 & 12.93 & 13.93 & 12.11 & 3.19 &  18.24 \\
    FD & AN-MVDR & 9.5M & 3.26 & 13.26 & 13.95 & 15.33 & 16.41 & 14.52 & \textbf{3.51} & 7.02\\
    FD & AN-MCWF & 9.4M & 3.25 & 12.78 & 13.66 & 15.19 & 16.13 & 14.26 & 3.49 & 7.40 \\
    \hline
    LD & eq-TI-MCWF & 9.2M & 7.51 & 3.78 & 4.84 & 6.97 & 8.25 & 5.71 & 2.52 & 51.40 \\
    LD & eq-TV-MCWF & 9.2M & 7.52 & 12.29 & 13.1 & 14.37 & 15.17 & 13.58 & 3.15 & 8.65 \\
    \hline
    TD & eq-MCWF & 9.2M & 7.52 & 13.92 & 14.47 & 15.56 & 16.32 & 14.94 & 3.29 &  8.72 \\
    TD & AN-MVDR & 10.4M & 11.18 & 15.10 & 15.66 & 16.46 & 17.07 & 15.97 & 3.30 &  6.74 \\
    TD & AN-MVDR (mch mask) & 11.3M & 11.75 & 15.56 & 16.23 & 17.31 & 17.92 & \textbf{16.63} & 3.43 & \textbf{6.20} \\
    TD & AN-MCWF & 10.3M & 8.48 & 14.73 & 15.27 & 16.19 & 16.86 & 15.65 & 3.31 & 7.12 \\
    TD & AN-MCWF (mch mask) & 10.8M & 10.02 & 15.35 & 15.96 & 16.96 & 17.61 & 16.35 & 3.39 & 6.61 \\
    \hline \hline
    FD & Ideal Binary Mask & - & - & 12.80 & 12.69 & 12.46 & 12.68 & 12.67 & 3.51 & 2.68 \\
    FD & Ideal Ratio Mask & - & - & 12.92 & 12.84 & 12.62 & 12.84 & 12.82 & 3.86 & 2.07 \\
    FD & Ideal Phase-Sensitive Mask & - & - & 16.96 & 16.85 & 16.64 & 16.83 & 16.83 & 4.05 & 2.05 \\
    FD & oracle MCWF & - & - & 12.88 & 13.46 & 15.97 & 17.26 & 14.62 & 3.34 & 14.87 \\
    TD & oracle MCWF & - & - & 49.84 & 50.26& 49.95 & 50.23 & 50.07 & 4.46 & 2.06 \\
    \hline \hline
  \end{tabular}
\end{table*}

\subsection{Performances of separation stage}

To quantify the effects of STFT- and learnable filter-based encoders, we conduct an ablation study for the designed features in frequency domain and time domain, respectively. The MC-SS results are summarized in Table \ref{tab:rlt_sep} in terms of SI-SDR, PESQ and CER. It should be noted that the CER of reverberant clean speech is 1.74\% and the state-of-the-art result on AISHELL-1 test set is about 4.1\% \cite{tian2022integrating,ren2022improving}. This is because the adopted ASR model is trained on 60 k hours data and our simulated test set did not include original full test set of AISHELL-1.

When only spectral feature is used (LPS v.s. R), the formulation of learnable filters achieves better single-channel speech separation performance over the STFT-based model, as demonstrated in \cite{luo2019convtasnet, bahmaninezhad2019comprehensive}. The introduction of spatial information in the frequency domain (+IPD) leads to SI-SDR gain of 0.65 dB. Extracting spatial features based on learnable filters (+ICD) or simply concatenating multichannel encodes (8-ch R) further improves the separation performances, which demonstrates the spatial feature formulation based on learnable filters may have the potential to better capture the spatial difference between speakers. When the target DOA is given, target speech separation can be performed via associating the output with the corresponding input direction. It can be observed that MC-SS models significantly benefit from the DOA information (+FD-DF or +LD-DF). Compared with the STFT based MC-SS model in frequency domain, the proposed learnable filter based MC-SS model in time domain improves the SI-SDR by 16.1\% and decreases CER by 13.7\%, which demonstrates a more powerful feature extraction capability of the learnable filters.
The high time resolution (short window length and hop size) of the features, as well as the formulations of learnable filters and features contributes to the performance improvement. 

It is also noted that, compared to single-channel Conv-TasNet (R), our proposed MC-SS model (R, ICD, TD-DF) only adds less than 0.1 million parameters, which come from the increased dimension of the layer that processes the input features. Also, the extra computation cost is relatively low since only a few convolution operations are involved.

\subsection{Performances of beamforming stage}
To evaluate the proposed AN-BF method in time domain, we compare the neural beamforming (eq-BF) and all-neural beamforming (AN-BF) based MC-SS models in frequency domain (FD) and time domain (TD). The input features are LPS+IPD+FD-DF for frequency domain methods, and R+ICD+LD-DF for time domain and latent domain methods. The SI-SDR, PESQ and CER results, as well as the number of parameters (\#param.) and multiply–accumulate operations per second (MACs), are listed in Table \ref{tab:rlt_bf}. We also compare two types of beamforming formulations: MVDR and MCWF, referred to AN-MVDR and AN-MCWF in the result table. Additionally, for the reference, we compute the performances achieved by ideal masks and oracle frequency domain and time domain beamformers. The ideal masks include ideal binary mask (IBM), ideal ratio mask (IRM) and ideal phase-sensitive mask (IPSM) \cite{wang2018supervised}. The oracle frequency domain MCWF is computed by Eq.\ref{eq:eq_mcwf}, where the ground truth target spectrogram $S$ is used to compute $\Phi_{ys}$. The oracle time domain MCWF is derived by Eq.\ref{eq:td_mcwf}, where the ground truth target speech $s$ is used to compute $\mathbf{r}_{ys}$.

Firstly, for frequency domain beamforming, FD-eq-MVDR and FD-eq-MCWF are performed using beamforming weights calculated by Eqs. \ref{eq:eq_mvdr} and \ref{eq:eq_mcwf}, respectively. Compared to masking based models, the performances of FD-eq-MVDR and FD-eq-MCWF are worse due to the time-invariant processing. 
The AN-BF method significantly improves the SI-SDR by 2.1 dB for FD-AN-MVDR \cite{xu2021generalized}. Also, the CER is relatively reduced by 30\%. This improvement benefits from the SCM modeling using recurrent network to automatically accumulate and update the spatial information.
Note that the filter based processing and multi-tap setup in \cite{zhang2020adl} are not adopted for fair comparison with time domain methods. The performance of FD-AN-MCWF is slightly worse than FD-AN-MVDR, since FD-AN-MVDR explicitly follows the target distortionless constraint. The extra AN-BF module brings about 5x MACs (0.6G v.s. 3.2G) due to the per-frequency processing.

Then, we conduct two neural beamforming experiments in latent domain (LD): time-invariant MCWF (TI-MCWF) and time-variant MCWF (TV-MCWF). The time-invariant beamforming weights are calculated by Eq. \ref{eq:eq_mcwf}, where $\Phi_{yy}(f)=\sum_{t}\mathbf{Y}(t, f')\mathbf{Y}^\mathsf{T}(t, f')$, $\Phi_{ys}(f)=\sum_{t}\mathbf{Y}(t, f')\hat{S}(t, f')$, $\mathbf{Y}(t, f')$ is the real-valued mixture representation, $\hat{S}(t, f')=\hat{M}_s(t, f') \circ \mathbf{Y}^\text{ref}(t, f')$ is the estimated target speech representation. The time-variant beamforming weights are derived using SCMs without summing over $t$. We can see these two formulations do not achieve better results compared to latent domain masking, since the beamforming derivation in latent domain remains unclear.

Next, for time domain beamforming, we compare neural beamforming (TD-eq-MCWF) and proposed all-neural beamforming (TD-AN-MVDR and TD-AN-MCWF) methods. The idea of time domain neural beamforming (TD-eq-MCWF) using the coarse estimate from the separation stage is similar to \cite{luo2022time} and achieves improvement of 0.5 dB SI-SDR. We extend the closed-form solution of MVDR and MCWF to all-neural formulations, and achieve SI-SDR gains of 1.57 dB and 1.15 dB, respectively. The CER reduces by 22.3\% and 18.0\% comparing to the separation stage, respectively. Also, compared to FD-AN-BF methods, our proposed TD-AN-MVDR outperforms FD-AN-MVDR by 2.1 dB of SI-SDR and reduces relatively 4\% of CER, while the proposed TD-AN-MCWF outperforms FD-AN-MCWF by 2.3 dB. The improvements demonstrate the effectiveness of our proposed AN-BF method in time domain. Furthermore, as discussed in Section \ref{subsec:mch_mask}, the proposed multichannel mask estimation strategy further improves TD-AN-MVDR and TD-AN-MCWF by 0.7 dB of SI-SDR and about 0.1 of PESQ. Also, the CER of TD-AN-MVDR (mch mask) is relatively reduced by 11.7\% compared to the best result achieved by FD-AN-MVDR. The best SI-SDR performance achieved by our proposed TD-AN-MVDR with multichannel mask estimation (16.63 dB) surpasses the IBM (by 4.0 dB), IRM (by 3.8 dB), oracle frequency domain MCWF (by 2.0 dB), and is comparable to the IPSM. Note that the PESQ scores of time domain methods fall behind frequency domain methods, because PESQ is evaluated only using the magnitude of short-time Bark-scale spectrum, which omits the phase \cite{wang2021compensation}. Also, the ASR model is not sensitive to the phase estimation error, which can be obviously observed when compared IRM (SI-SDR of 12.82 dB and CER of 2.07\%) to IPSM (SI-SDR of 16.83 dB and CER of 2.05\%). For magnitude enhancement purpose, the PESQ and CER can be improved by alternating the loss function from SI-SDR to magnitude based logarithm MSE \cite{wang2021compensation}. 

Because of the short hop size, the MACs of all of the time domain methods is relatively high. Compared to FD-AN-BF methods, the proposed TD-AN-BF module only brings about 15\%—50\% extra MACs (7.5G v.s. 8.5—11.5G). 

\begin{table}[t]
\small
  \caption{SI-SDR (dB), PESQ and CER (\%) results of FD-AN-MVDR and TD-AN-MVDR models that finetuned (FT.)  with L-MFB MSE loss or not.}
  \label{tab:rlt_ft}
  \centering
  \begin{tabular}{l|c|c|c|c}
    \hline
    \hline
    \textbf{Approach} & \textbf{FT.} &\textbf{SI-SDR (dB)} & \textbf{PESQ} & \textbf{CER (\%)} \\ 
    \hline
    FD-AN-MVDR & \XSolidBrush & 14.52 & 3.51 & 7.02 \\
    FD-AN-MVDR & \Checkmark & 14.29 & 3.53 & 6.81 \\
    \hline
    TD-AN-MVDR & \XSolidBrush & 16.63 & 3.43 & 6.20  \\
    TD-AN-MVDR & \Checkmark & 16.42 & 3.49 & 5.67 \\
    \hline \hline
  \end{tabular}
\end{table}

\begin{table}[t]
  \caption{SI-SDR (dB), PESQ and CER (\%) results of MC-SS models in frequency domain and time domain, with different window sizes $N$ and hop sizes $H$.}
  \label{tab:rlt_hop}
  
  \small
  \centering
  \begin{tabular}{l|c|c|c}
    \hline
    \hline
    \textbf{Approach} & \textbf{SI-SDR (dB)} & \textbf{PESQ} & \textbf{CER (\%)} \\ 
    \hline \hline
    \multicolumn{4}{c}{\textbf{Frequency Domain} ($N$=512, $H$=256, nFFT=512)} \\
    \hline 
    cRM &  12.40 & 3.23 & 10.06 \\
    AN-MVDR  & 14.52 & 3.51 & 7.02 \\
    AN-MCWF  & 14.26 & 3.49 &  7.40 \\
    \hline \hline
    \multicolumn{4}{c}{\textbf{Frequency Domain} ($N$=40, $H$=20, nFFT=64)} \\
    \hline 
    cRM  &  11.04 & 2.95 & 12.86 \\ 
    AN-MVDR  & 9.61 & 2.83 & 27.32 \\
    AN-MCWF  & 9.49 & 2.82 & 28.05 \\
    \hline\hline
    \multicolumn{4}{c}{\textbf{Time Domain} ($N$=40, $H$=20)} \\
    \hline
    Mask & 14.40 & 3.26 & 8.68 \\
    AN-MVDR & 15.97 & 3.30  & 6.74 \\
    AN-MVDR (mch mask)  & 16.63 & 3.43 & 6.20\\
    AN-MCWF  & 15.65 & 3.31 & 7.12 \\
    AN-MCWF (mch mask) & 16.35 & 3.39 & 6.61\\
    \hline \hline
    \multicolumn{4}{c}{\textbf{Time Domain} ($N$=512, $H$=256)} \\
    \hline
    Mask &  9.00 & 2.65 & 22.53 \\
    AN-MVDR & 10.50 & 2.69 & 16.93 \\
    AN-MCWF  & 10.34 & 2.67 & 17.19 \\
    \hline\hline
  \end{tabular}
\end{table}

\begin{figure*}[h]
\centerline{\includegraphics[width=15cm]{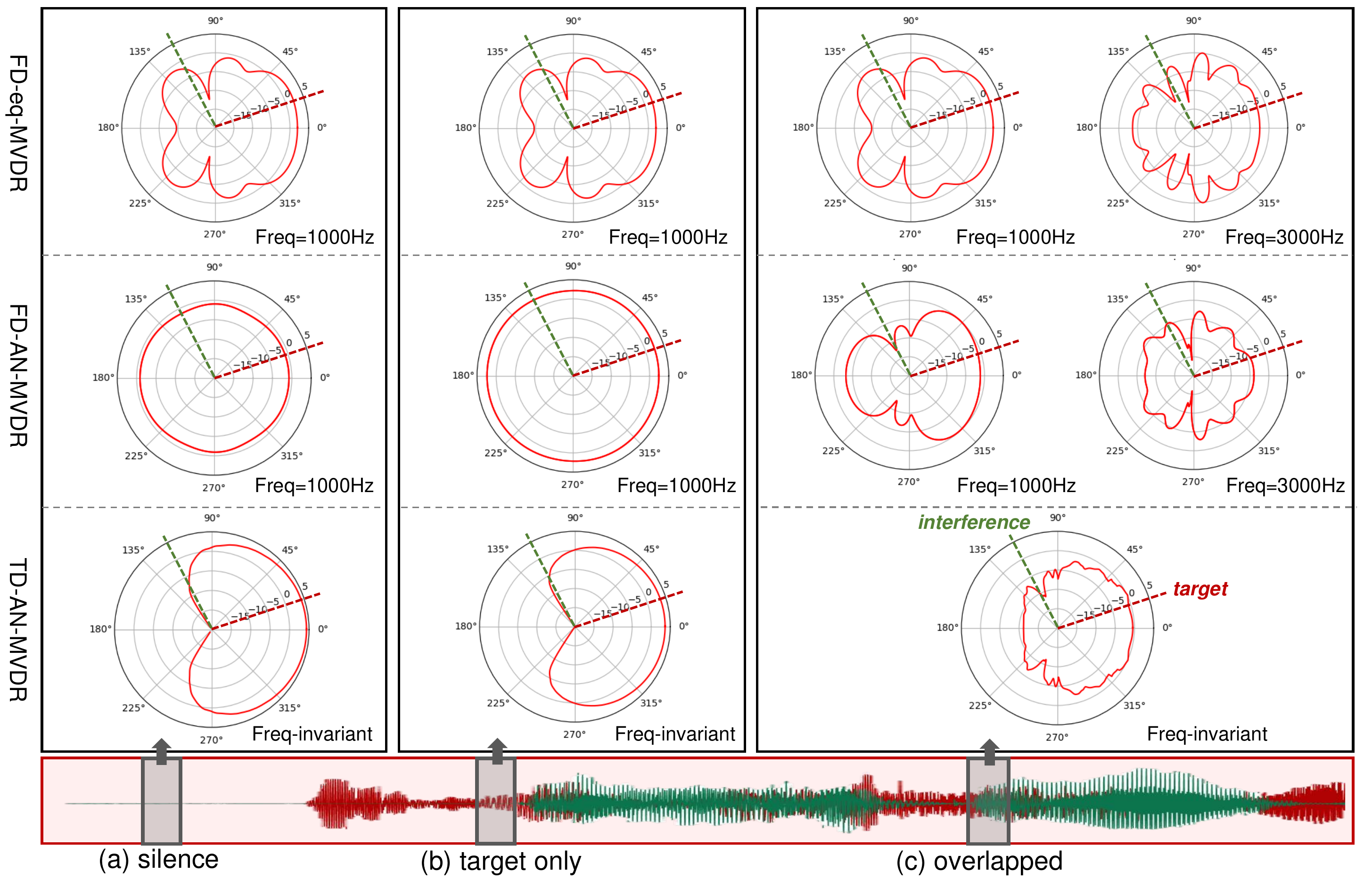}}
\caption{An example of spatial beam patterns of beamformers derived by FD-eq-MVDR, FD-AN-MVDR and TD-AN-MVDR models. The speech sample contains two speakers. The DOAs of the target and interference speaker are 22$\degree$ and 120$\degree$, respectively marked with red and green dotted lines.}
\label{fig:beampattern}
\end{figure*}
\subsection{Finetuning with logarithm Mel-scaled filterbank MSE loss}
To demonstrate that the PESQ and CER metrics of TD-AN-MVDR can be improved by alternating with or adding a magnitude-based loss, we finetune the FD-AN-MVDR and the proposed TD-AN-MVDR models with a logarithm Mel-scaled filterbank (L-MFB) MSE loss:
\begin{equation}
L_\text{L-MFB}=\left | \log ( (\text{MFB}(\hat{S}_\text{BF}) ) - \log\left(\text{MFB}(S) \right) \right |^2 
\label{eq:lfb_mse}
\end{equation}
where $\hat{S}_\text{BF}$ and $S$ are spectrograms of the beamformed signal and the target signal, $\text{MFB}$ means to apply the Mel-scaled filter bank to the spectrogram. The converged model trained with SI-SDR loss is finetuned for extra 20 epochs using a initial learning rate of 1e-4. The learning rate will be halved if the validation loss does not improve for consecutive 3 epochs. 

The updated performances are reported in Table  \ref{tab:rlt_ft}. After the magnitude enhancement, the PESQ and CER results of TD-AN-MVDR model achieves a more obvious improvement over those of FD-AN-MVDR model. The CER drops from 6.20\% to 5.67\%, which further narrows the gap between the proposed method and the reference speech.   

\subsection{Effects of window and hop size}
As pointed by \cite{heitkaemper2020demystifying}, the success of Conv-TasNet may benefit from the high time resolution, i.e., small window size and hop size, which is important to isolate feature bins dominated by different speech. Therefore, we compare the AN-BF methods in FD and TD using different window sizes and hop sizes. 

In Table \ref{tab:rlt_hop}, we evaluate two setups of FD methods. One is the common setup, where the window size is 32 ms ($N$=512), hop size is 16 ms ($H$=256), and the FFT size is 512. The other setup refers to that of TD method, where the window size is set as 2.5 ms ($N$=40) and the hop size is 1.25 ms ($H$=20). The FFT size is set to the nearest exponential of 2 to 40, i.e., 64. Compared these two setups, we can see that the shorten of window size does not improve but deteriorate the separation performance. The reason maybe that, when the window and hop sizes are short, the corresponding frequency resolution decreases as well, which is adverse to the separation process. 

When applying the same window and hop size setups to TD methods, contrary separation performances are observed. TD methods prefer short window and hop sizes, while large hop size severely hurts the performance. This observation is also reported in \cite{kolbaek2021tasnet}, probably caused by the lack of anti-alising filters to deal with the larger aliasing components brought by longer hop size. 

\begin{table}[b]
  \caption{Performance comparisons with advancing MC-SS methods.}
  \label{tab:rlt_other}  
  
  \centering
  \scalebox{0.75}{
  \begin{tabular}{l|c|c|c|c|c}
    \hline
    \hline
    \textbf{Approach} & \#\textbf{param.} & \textbf{MACs} &\textbf{SI-SDR (dB)} & \textbf{PESQ} & \textbf{CER (\%)} \\ 
    \hline \hline
    NB-BLSTM \cite{quan2022multi} & 1.5M & 19.85G& 12.78 & 3.35 & 8.88 \\
    \quad + FD-DF* & 1.5M & 19.91G & 8.29 & 2.87 & 25.98\\
    \hline
    FaSNet-TAC \cite{luo2020end} & 2.6M & 19.44G & 12.94 & 2.93 & 13.56 \\
    \quad + LD-DF* & 2.6M & 19.46G & 13.17 & 2.97 & 11.25 \\       
    \hline \hline
    TD-AN-MVDR  & 10.4M & 11.18G & 15.97 & 3.30 & 6.74 \\    
    \hline\hline
  \end{tabular}}
\end{table}

\subsection{Comparison with advanced methods}
This subsection compares the proposed method with some advanced multichannel speech separation or enhancement methods. Generally, these methods can be categorized into frequency domain and end-to-end (time domain) methods. FD methods basically follow the T-F masking based (neural) beamforming scheme, while varying in the design of temporal-spectral-spatial information processing \cite{halimeh2022complex,ren2021causal, quan22b_interspeech, quan2022multi}, beamforming derivation \cite{li2022taylorbeamformer} and multi-stage filtering scheme \cite{tesch2021nonlinear,tesch2022insights}. 
End-to-end methods perform signal estimation and beamforming in latent domain or time domain in an end-to-end manner \cite{luo2020end,luo2021implicit,luo2022time}.

In Table \ref{tab:rlt_other}, we repeat two advanced blind speech separation methods for comparison: \emph{NB-BLSTM} \cite{quan2022multi} in frequency domain and \emph{FaSNet-TAC} \cite{luo2020end} in time domain. To make fair comparison, we also experiment with concatenating the DOA features (FD-DF, LD-DF) with the input features as additional target DOA cues to perform target speech separation. It is noted that our implementation may not be the best way to integrate DOA cues into these architectures. Therefore, improved performance can be promising with more elaborate design. 

NB-BLSTM performs blind speech separation independently at each subband, where the narrow-band processing concept is similar to AN-BF \cite{zhang2020adl}. NB-BLSTM exhibits impressive PESQ and CER results with fewer parameters. However, the subband processing leads to high computation cost. Also, although FD-DF informs the NB-BLSTM model of the target DOA information, the separation performance is degraded. This may be caused by cross-frequency discontinuity of the target information.
FaSNet-TAC performs transform-and-concatenate operation within each DPRNN block \cite{luo2020dual} to integrate multichannel information, where DPRNN is a state-of-the-art architecture for speech separation. LD-DF enables FaSNet-TAC to perform target speech separation and also slightly improves the performance, which demonstrates our proposed directional feature can be integrated into time domain separation architectures for further advance.

\subsection{Comparison of beam patterns}

In order to further analyze the model capability of enhancing the signal from target direction while suppressing interference from other directions, Figure \ref{fig:beampattern} visualizes the spatial beam patterns of beamformers learned by different models, including the time-invariant MVDR beamformer in frequency domain (FD-eq-MVDR), the time-variant beamformer in frequency domain (FD-AN-MVDR) and the proposed time-variant beamformer in time domain (TD-AN-MVDR). We visualize the beam patterns under 3 kinds of scenarios, including (a) both speakers are silent; (b) only the target speaker is active and (c) both speakers are simultaneously talking. 

For FD-eq-MVDR, the beam patterns are time-invariant and frequency-dependent. The gain at interference direction (120$\degree$) is relatively low to suppress the interfering speech. However, the side lobe leakage may be serious. For FD-AN-MVDR, the beam patterns of FD-AN-MVDR are time-variant and frequency-dependent. We can observe an interesting phenomenon that the beam pattern during silence and single-speaker speech regions has no directivity. This can be good, which better preserves the target signal without any distortion. Although the beam pattern exhibits fair directivity at lower frequency (1000 Hz), it is not desirable at higher frequency (3000 Hz). The null points to nearly 90$\degree$ rather than 120$\degree$, which brings potential distortion to the target signal at higher frequencies. For TD-AN-MVDR, the beam patterns are frequency-invariant. In three cases, the beamformer show desirable patterns, which tends to suppress the interfering signal from the range of 120$\degree$ to 240$\degree$ and enhance the signal from opposite directions.

\section{Conclusion}
\label{sec:conclusion}

In this work, we develops a fully time domain all-neural beamforming network, which features with temporal-spectral-spatial feature learning and parametric beamforming coefficient estimation in time domain for better target signal estimation. The proposed feature learning method can be integrated into popular time domain network architectures, with few parameters and computation cost increased, to enable target speech separation. Also, the all-neural beamforming method can be combined with state-of-the-art mask estimation methods to further enhance the separation accuracy. Experimental results demonstrate that the proposed methods achieve good results in terms of objective evaluation metrics and ASR accuracy, and outperform frequency domain state-of-the-arts as well as time domain neural beamforming methods. 

In addition, we make an attempt to unify the all-neural beamforming pipelines for time domain and frequency domain multichannel speech separation. The waveform encoder, feature formulation, mask estimation, as well as the all-neural beamforming designs in time and frequency domain are described, analyzed and evaluated in a comparable way. 

We will work in the following directions to extend this study in the future. Firstly, as shown in Table \ref{tab:rlt_bf} and \ref{tab:rlt_ft}, the PESQ results of time domain methods still fall behind frequency domain methods. We will further explore approaches to enhance the magnitude part estimation, which will also be beneficial for the popular ASR models that receive logarithm MFB or MFCC features as input. Secondly, we will evaluate the framework on more large-scale and realistic datasets, with different overlap ratios, more speakers, etc. Lastly, we will extend the all-neural beamforming framework for joint separation and dereverberation tasks. 

\bibliographystyle{IEEEtran}
\bibliography{main}

\end{document}